# Interfacial bound exciton state in a hybrid structure of monolayer $WS_2$ and InGaN quantum dots

Guanghui Cheng[a], Baikui Li[a,b,1], Chunyu Zhao[a,d], Xin Yan[a], Hong Wang[c], Kei May Lau[d], and Jiannong Wang[a,1]

[a]Department of Physics, Hong Kong University of Science and Technology, Clear Water Bay, Kowloon, Hong Kong

[b]College of Optoelectronic Engineering, Shenzhen University, Nanhai Ave 3688, Shenzhen, China

[c]Department of Materials Science and Engineering, Southern University of Science and Technology, Shenzhen 518055, China

[d]Department of Electronic and Computer Engineering, Hong Kong University of Science and Technology, Clear Water Bay, Kowloon, Hong Kong

[1]To whom correspondence should be addressed:
Baikui Li, Jiannong Wang
Emails: libk@szu.edu.cn, phjwang@ust.hk

**Abstract**

Van der Waals heterostructures usually formed using atomic thin transition metal dichalcogenides (TMDCs) with a direct bandgap in the near-infrared to the visible range are promising candidates for low-dimensional optoelectronic applications. The interlayer interaction or coupling between two-dimensional (2D) layer and the substrate or between adjacent 2D layers plays an important role in modifying properties of the individual 2D material or device performances through Coulomb interaction or forming interlayer excitons. Here, we report the realization of quasi zero-dimensional (0D) photon emission of $WS_2$ in a coupled hybrid structure of monolayer $WS_2$ and InGaN quantum dots (QDs). An interfacial bound exciton, i.e., coupling between excitons in $WS_2$ and the electrons in QDs has been identified. The emission of this interfacial bound exciton inherits the 0-D confinement of QDs as well as the spin-valley physics of excitons in monolayer $WS_2$. The effective coupling between 2D materials and conventional semiconductors observed in this work provides an effective way to realize 0D emission of 2D materials and opens the potential of compact on-chip integration of valleytronics and conventional electronics/optoelectronics.

*Keywords: transition metal dichalcogenides, quantum dots, interfacial state, photoluminescence*

**Significance statement**

The van der Waals heterostructure of low dimensional materials offers new opportunities to manipulate fascinating optoelectronic phenomena for device applications. The interlayer coupling between the constituents is greatly enhanced due to the sharp interface and changing screening effect. In this paper, we observe a novel radiative state in the hybrid structure combining monolayer $WS_2$ with InGaN quantum dots (QDs), which is identified to be the Coulomb coupling between excitons in $WS_2$ and electrons in QDs. This interfacial bound exciton emission inherits the zero-dimensional (0D) confinement of QDs as well as the spin-valley physics of excitons in $WS_2$, providing rich insight into realizing 0D emission of two-dimensional materials and integrating valleytronics with conventional optoelectronics.

**Introduction**

Monolayer (ML) transition metal dichalcogenides (TMDCs) in the form of $MX_2$ (M=Mo, W; X=S, Se) show direct band gap with band extrema located at the finite momentum K and K' points of the hexagonal Brillouin zone and interband transitions in the visible to near-infrared spectral range (1). Due to the quantum confinement in two-dimensional (2D) MLs, and weak dielectric screening the exciton binding energy in TMDC MLs is of hundreds of meV (2, 3), which is one or two orders of magnitude higher than that of traditional semiconductors such as GaAs or GaN quantum wells used in today's optoelectronic devices (4). Moreover, the strong spin-orbit interaction and lack of inversion symmetry of TMDC MLs give rise to inequivalent spin splitting in the K and K' valleys, leading to the valley-dependent optical selection rule (5, 6). Therefore, TMDC MLs have great potential in high efficient optoelectronics (7, 8) as well as valleytronics (9). Recently, photodetectors and photodiodes based on the TMDC ML van der Waals (vdW) heterostructures, formed in a layer-by-layer assembly with a designed sequence, have been demonstrated (10-12). Due to the low-level disorder and changing screening effect within the atomically sharp interface, the interlayer coupling in vdW heterostructures is greatly enhanced through Coulomb interaction. This offers new opportunities to design and explore novel optoelectronic phenomena in vdW heterostructures (13). One example is the observation and modulation of the interlayer excitons, formed by binding the spatially separated electrons and holes under the staggered band alignment, in TMDC ML vdW heterostructures (14, 15). The manipulation of the valley degree of freedom of interlayer exciton has also been observed in TMDC ML vdW heterostructures (16, 17).

Furthermore, the dangling-bond free surface of TMDC MLs enables their easy integration with other low-dimensional materials forming mixed dimensional heterostructures, which can combine and utilize the respective functionalities (18). Semiconductor quantum dots (QDs) are optically active allowing for the realization of single-photon emission and spin-photon manipulation at the nanometer length scale (19, 20). Therefore, bringing QDs and TMDC MLs together would allow us to utilize their respective strengths, understand their interfacial interactions, and explore new properties for potential applications in optoelectronics and valleytronics. Earlier studies have shown photoluminescence (PL) modulation and energy transfer depending on the device structures and the band alignment (21-23). However, the essential Coulomb couplings between the two quantum paradigms and the recombination dynamics have not yet been investigated in details. In this study, we have fabricated coupled hybrid structures of monolayer $WS_2$ and self-assembled InGaN QDs. In photoluminescence studies, we observe a remarkable emission with energy slightly below the intralayer free exciton of $WS_2$. This emission is identified to originate from the interfacial bound excitons formed by coupling excitons in $WS_2$ and electrons in QDs. In addition, the time-resolved, the excitation power and polarization dependent PL studies reveal that this interfacial bound exciton emission inherits the zero-dimensional (0D) confinement of QDs as well as the spin-valley physics of excitons in $WS_2$.

## Results and Discussion

Hybrid structures are prepared using dry-transfer techniques by integrating the exfoliated $WS_2$ monolayer (ML), the thick boron nitride (BN) slab (~ 90 nm) and the MOCVD-grown InGaN QDs on sapphire (see Methods). The schematic sample configuration and optical microscope image of a typical device are shown in Fig. 1*A* and 1*B*, respectively. The device consists region-I with bare InGaN QDs, region-II the uncoupled $WS_2$/BN/QDs stack, and region-III the coupled $WS_2$/QDs stack. Such a device structure allows for the investigation of the interfacial coupling between 2D $WS_2$ ML and quasi-0D InGaN QDs by comparing the optical transition states in three regions. The 400 nm excitation laser beam is primarily used in this work, where the photon energy is large enough to excite the interband transitions in both $WS_2$ ML and InGaN QDs. All of our steady-state and time-resolved PL measurements are performed at 4 K unless otherwise stated (see Methods). The PL spectrum of InGaN QDs in region-I is shown in the inset of Fig. 1*C* upper panel. The emission peak is located at 2.38 eV, which is attributed to the allowed lowest exciton state of InGaN QDs. No radiative recombination from InGaN wetting layer is observed due to the fast relaxation to the low-lying QD levels (24).

Figure 1c shows the PL spectra of $WS_2$ ML from the uncoupled region-II (upper panel) and the coupled region-III (lower panel). As it can be seen, two measured PL spectra (solid symbols) exhibit double-peak line-shape which can be well fitted (solid and dashed lines) with Voigt function, i.e., a convolution of a Lorentzian and a Gaussian representing the influence of homogeneous and inhomogeneous broadening (25). The Voigt fittings reveal that for the uncoupled region-II PL spectrum is dominant with the peak at 2.036 eV, corresponding to the well-known bright band-edge exciton emission (denoted as A) at K and K' valleys of $WS_2$ ML (26). While a shoulder at 2.000 eV represents the negative trion emission (denoted as T) formed by binding an exciton with an intervalley or intravalley free electron, as the exfoliated $WS_2$ is in general n-type (27). The binding energy for the trion is thus extracted to be 36 meV, dominated by the Coulomb interaction between the exciton and the free electron (28). On the other hand, for the coupled region-III, the exciton A peak slightly blue shifts to 2.037 eV and the trion T peak disappears, while a new prominent peak emerges at 2.019 eV (denoted as X) with its intensity even stronger than that of the exciton A peak. Fig. 1*D* is the spatial mapping of PL emissions at A, T, X, and QDs peak energies for the hybrid structure shown in Fig. 1*B*. It is interesting to note that T emission only locates in the uncoupled region II which implies that the coupling of WS$_2$ ML and QDs results in free electron transfer from WS$_2$ to QDs. In contrast, X emission locates in the coupled region III indicating it is the coupling induced emission. In addition, the total PL emission intensity of WS$_2$ ML is greatly enhanced in the coupled region III while the PL intensity of QDs in the same region is reduced.

These results can be well understood from the band structure of the coupled WS$_2$ ML and InGaN QDs hybrid structure. The In composition of our InGaN QDs is deduced to be 28% with estimated electron affinity of 4.82 eV and the band gap energy of 2.43 eV (*SI Text*, section 1). While the electron affinity of WS$_2$ ML is known to be 3.75 eV (29, 30) and the direct band gap energy is 2.75 eV. These give rise to a type-II band alignment across WS$_2$ ML and InGaN QDs as schematically shown in Fig. 2*A*, where both of the conduction band minimum and valence band maximum of InGaN QDs are below that of WS$_2$ ML. As a result, in the coupled region III free electrons in n-type WS$_2$ readily transfer to InGaN QDs leading to a quenching T emission in PL spectra. The band bending of InGaN QDs in the band diagram is caused by the spontaneous and piezoelectric polarization-induced electric field (31-33). This band bending causes the photoexcited electrons in InGaN QDs to accumulate at the interface, which can form a lateral potential well and easily capture the A excitons in WS$_2$ ML to form an interfacial negatively charge exciton, X. The Coulomb interaction in this case is expected to be weaker than that for intralayer trions within WS$_2$ ML so as to produce an X emission with peak energy larger than T but smaller than A, as we have observed (see Fig. 1*C*). The capturing or lateral localization of excitons in WS$_2$ ML by QDs, i.e., quasi-0D bound excitons, reduce their in-plane diffusion and in-turn non-radiative recombination rate giving rise to an enhanced overall PL emission intensity.

In order to further verify the above proposed band alignment and clarify the origin of X emission, PL spectra from the coupled region III have been measured with laser excitation wavelength at 593 nm (see Fig. 2*B*). At this photon energy only excitons in WS$_2$ ML can be excited. As it can be seen the PL spectra show only A exciton emission from WS$_2$ ML and no X emission is detectable. This result indicates that X emission is indeed associated with photoexcitation of InGaN QDs, where photoexcited electron accumulation at the interface is believed to couple with A excitons in WS$_2$ ML forming interfacial charged exciton X. Furthermore, a hybrid structure consisting of WS$_2$ ML and InGaN QDs separated by ML BN has been prepared. With ML BN insertion at the interface the coupling between WS$_2$ and InGaN QDs is slightly weakened and the free electron transfer from WS$_2$ to InGaN QDs is partially blocked. As a result, we expect to observe WS$_2$ T emission in addition to A and X emissions if both WS$_2$ ML and InGaN QDs are photoexcited. Remarkably, as it is shown in Fig. 2*C*, the PL spectra measured with laser excitation wavelength at 400 nm indeed consists of A, T, and X emissions. The slight weakening of coupling with ML BN insertion is evidenced by the slight decrease of the binding energy of X to 17 meV.

To study the dynamic behaviors of the interfacial negatively charged exciton X, as well as the effect of interlayer coupling on that of exciton A, time-resolved PL (TRPL) measurements of the hybrid structure have been carried out. The TRPL images, time evolutional PL spectra, and the decay curves at each peak wavelength, measured from uncoupled region-II and coupled region-III are shown in Fig. 3*A-C* and 3*D-F*, respectively. In the uncoupled case, the luminescence lifetimes of A and T are extracted to be 39 and 48 ps, respectively, by fitting the decay curves in Fig. 3*C* using a single exponential equation. In the coupled case, the decay of A and X emission also follows a single exponential curve, however, with a lifetime of 33 and 55 ps, respectively. As the InGaN QDs PL emission lifetime is a few nano-seconds (*SI Text*, section 2 and Fig. S1), the photoexcited electron-hole pairs density in QDs can be considered unchanged within the time scale of $WS_2$ ML PL emissions. The above results show that the emission of interfacial negatively charged exciton X has a lifetime marginally larger than that of the intralayer negatively charged exciton T. This is probably due to the smaller binding energy of X than that of the T. In the coupled region, exciton A locates on the surface of InGaN wetting layer (WL). The surface charge of WL or the polarization charge at the WL/GaN interface will produce a screening effect on exciton A, consequently resulting in the observed blue shift and the slightly smaller decay time constant of A emission compared with that in the uncoupled region. In addition, the formation of T or the coupling between excitons and electrons in uncoupled case is an ultrafast process, as revealed by the simultaneous rising of luminescence intensities of A and T at almost the same rate shown in Fig. 3*C*. However, in the coupled case, the rising of X is obviously slower than that of A and has a delay of ~7.4 ps to reach its maximum intensity, as shown in Fig. 3*F*. This delay is not due to the slower formation time of X since the luminescence intensities of A and X start to rise simultaneously, but is due to the time required for A to diffuse laterally towards the spatially distributed QDs where the electrons with high density are located.

Two excitation-power-dependent phenomena of $WS_2$ ML have been reported (34, 35). With higher excitation power, (i) the relative intensity of T emission increases because of the higher photon-excited-electron density and the consequent higher formation probability of T, and (ii) the peak position of A emission shows a blue shift due to the state filling effect. The PL behaviors of $WS_2$ ML in the uncoupled region II with different excitation powers have been studied. As expected, the relative intensity of T emission increases and the peak position of A emission shows a blue shift with increasing excitation power, as shown in Fig. 4*A*. However, opposite excitation-power dependences of PL are observed in the coupled region III. As shown in Fig. 4*B*, the relatively intensity of X decreases and the peak positions of A and X emissions show a red shift with increasing excitation power. The decrease of relatively intensity of X emission can be understood by considering the spatial distribution of the QDs or the localization of electrons. Although the electron density in the QDs increases at higher excitation powers, the number of A can be coupled with the localized electrons in a QD are finite due to limited allowed spatial density. Therefore, at higher excitation power density, the portion of A that can diffuse to the QDs and be coupled with electrons therein to form X decreases, leading to the decrease of relatively intensity of X emission. Note that the increase of electron density in the QDs is evidenced by the blue shift of QD emission peak with increasing excitation power (*SI Text*, section 2 and Fig. S1).

As excitation power increases, both emission energies of X and A show red shifts. The peak energy of A emission in the coupled case is not only affected vertically by the charges in InGaN WL, i.e., screening effect which induces a blue shift, but also laterally by the electrons in QDs, i.e., Stark effect which induces a red shift. At low excitation powers, vertical screening effect dominates and the peak energy of A emission shows a blue shift of ~ 1 meV with little excitation-power dependence compared with that of the uncoupled region, while the electrons in QDs tend to couple with excitons to form X. At higher excitation powers, as the formation of X saturates as discussed above, the extra electrons in QDs will start to affect exciton A significantly through a Stark effect, consequently leading to a decrease of emission energy, i.e., a pronounced red shift as excitation power further increases. Rigorous inspection of the power dependences shown in Fig. 4*B* reveals the correspondence between the significant decrease of relatively intensity of X (formation saturation) and the significant red shift of A emission (threshold of lateral Stark effect) as excitation power increases, as marked by a vertical dashed line. On the other hand, we note that as excitation power increases, the red shift of emission energies of X is much larger than that of A. This additional red shift of X can be explained as following: At higher excitation power density, the electron density in QD increases while the number of capture A excitons cannot increase accordingly as discussed above. This results in a relatively stronger coupling between A excitons and the localized electrons, i.e., larger binding energy of X, and consequently leading to a further red shift of X emission.

The excitons in WS$_2$ ML carry the circular dichroism as a result of valley dependent optical selection rules. To study the polarization properties of the interfacial bound exciton X, circular polarization dependent PL were carried out on a coupled structure. As shown in Fig. 4*C*, the circular polarization defined as ($I_{LL}$ - $I_{LR}$)/($I_{LL}$ + $I_{LR}$) is the same (~ 16%) for A and X emission, where $I_{LL(LR)}$ is the intensity of PL with the left-hand-circularly-polarized-excitation/left-hand-circularly-polarized-detection (left-hand-circularly-polarized-excitation/right-hand-circularly-polarized-detection). This indicates that the X exciton inherits the spin-valley physics of A exciton, indicating the potentiality of this quasi zero-dimensional interfacial bound excitons in future valleytronics.

## Conclusions

In conclusion, we have observed a novel radiative state in the WS$_2$/QDs hybrid structure, which is identified to be the coupling between excitons in WS$_2$ and electrons in QDs through Coulomb interactions. The spatial localization of such interfacial bound exciton offers an effective way to realize 0D emission of 2D materials which is of significance in developing TMDC quantum photon emitters. Its robust luminescence and the spin-valley physics of this intercial bound exciton may lead to novel optoelectronic applications by integrating the 2D TMDC valleytronics with conventional electronics and optoelectronics.

## Materials and Methods

**InGaN QDs growth.** The InGaN QDs samples were grown in an Aixtron Close Coupled Showerhead (CCS) MOCVD reactor. Trimethylgallium (TMGa), trimethylindium (TMIn) and ammonia (NH$_3$) were employed as sources for Ga, In and elemental N, respectively. A thermal cleaning was performed to desorb the residual native oxides at the surface of the c-plane sapphire substrate, followed by a GaN nucleation layer, an unintentionally doped GaN layer and a 3-μm silicon-doped GaN layer to form a GaN pseudo-substrate. Prior to the growth of InGaN QDs, a 600-nm silicon-doped GaN connecting layer was grown to bury the regrowth interface and provide a smooth surface. An uncapped layer of InGaN QDs was grown at 670 °C with a V/III ratio of 1.02×10$^4$. The self-assembled quantum dots are formed by Stranski-Krastanov (SK) growth mode. The structure characterizations of InGaN QDs can be found in *SI Text*, section 3 and Fig. S2.

**Device fabrications.** Boron Nitride (BN) slab was mechanically exfoliated onto a thin layer of polydimethylsiloxane (PDMS), with thickness confirmed by the optical contrast and the AFM. The monolayer WS$_2$ flake was exfoliated onto another piece of PDMS, confirmed by optical contrast, PL spectra and AFM. By the aid of an optical microscope and micromanipulators, the BN and the WS$_2$ were then sequentially placed in contact with the QDs to form the hybrid structure depicted in the main text. Sample transfer processes were carried out in a nitrogen-filled glove box and then the device was rapidly mounted into cryostat optical chamber.

**Low-temperature steady-state and time-resolved PL measurements.** Low-temperature measurements were conducted in an Oxford microscopy cryostat with sample in vacuum. The laser beam were guided into an Olympus microscope and focused onto a specific position of the sample using a long-working-distance objective (Mitutoyo APO 100X), with spot diameter ~ 3 μm. The PL was detected by a high-resolution spectrometer. Spatial PL mapping was performed using Newport precision motorized actuators and a grating spectrometer SPEX 500M with a photomultiplier tube. For the time-resolved PL measurement, the sample was excited by frequency-doubled Ti: sapphire laser system with 200 fs pulses, 76 MHz repetition rate and 400 nm output wavelength. The PL signals were detected by a streak camera system (Hamamatsu) with picosecond resolution.

## Acknowledgements

We thank Prof. Steven G. Louie of the University of California at Berkeley for helpful discussions. This work is supported by the Research Grants Council of the Hong Kong SAR under Grant Nos. N_HKUST605/16 and C7036-17W, and also in part by Shenzhen Science and Technology Innovation Commission under Grant No. JCYJ20170412110137562 and in part by the National Natural Science Foundation of China (No.: 61631166004).

.

# Figure Legends

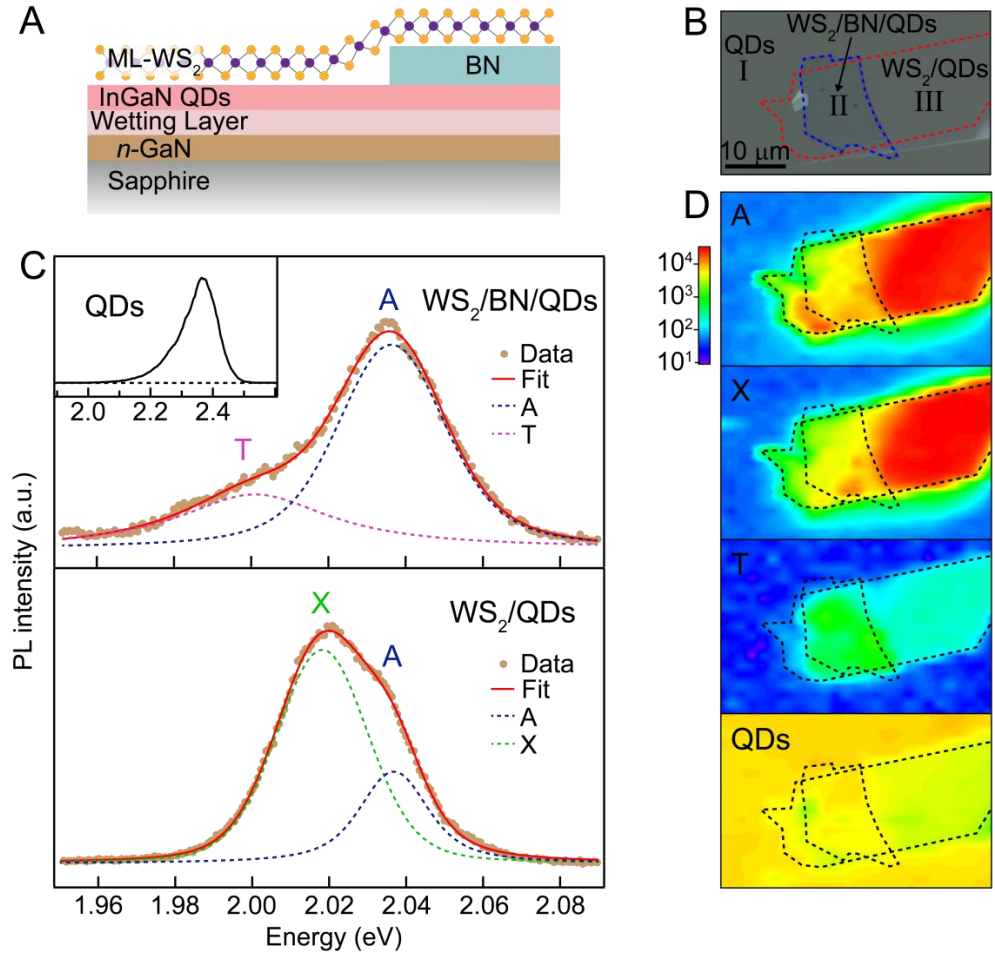

**Fig. 1.** PL spectra and spatial mappings of a WS$_2$-QDs hybrid structure. (*A*) Schematic illustration of the cross-section of a monolayer WS$_2$-InGaN QDs hybrid structure. (*B*) Image of the WS$_2$/QDs device. A red dashed line outlines the exfoliated WS$_2$, and a blue dashed line outlines an inserted thick BN slab. Region-I, II and III represent bare QDs, QDs/BN/WS$_2$ and QDs/WS$_2$ configurations, respectively. (*C*) The PL spectra from the uncoupled region-II (upper panel) and the coupled region-III (lower panel) with an excitation laser of 400 nm and 16 μW. The solid and dashed lines are fitting curves using Voigt functions. In region-II, the PL spectrum consists of emissions from exciton A and the trion T. In region-III, the PL spectrum consists of emissions from exciton A and a new state X. Inset: PL spectrum of the InGaN QDs. (*D*) Spatial mappings of the PL emission intensities from A, T, X and QDs as labelled. (It is slightly red shifted when mapping the T emission to avoid spectral overlap). PL intensity is denoted by the color according to the scale bar. Mappings are performed with a laser excitation at 400 nm, 2 μW.

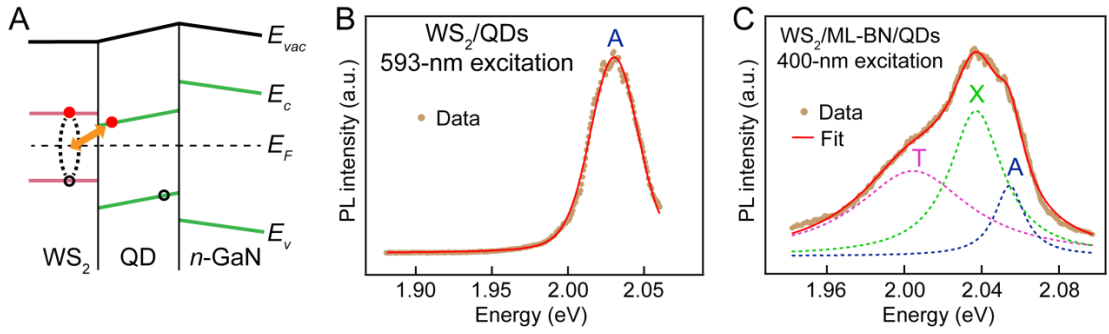

**Fig. 2.** Band diagram and origin of X emission. (*A*) A schematic band diagram of the coupled heterostructure of WS$_2$/QD/GaN. Interfacial binding between an exciton in WS$_2$ and an electron in QD is illustrated. (*B*) PL spectrum from the coupled region-III WS$_2$/QDs under 593-nm laser excitation, (solid line is a fitting curve using Voigt function), in which case only WS$_2$ is selectively excited. Only the emission peak of exciton A can be identified. (*C*) PL spectrum from a weakly coupled heterostructure WS$_2$/ML-BN/QDs under 400-nm laser excitation. The spectrum consists of emissions from A, T, and X. the solid lines are fitting curves using Lorentzian functions.

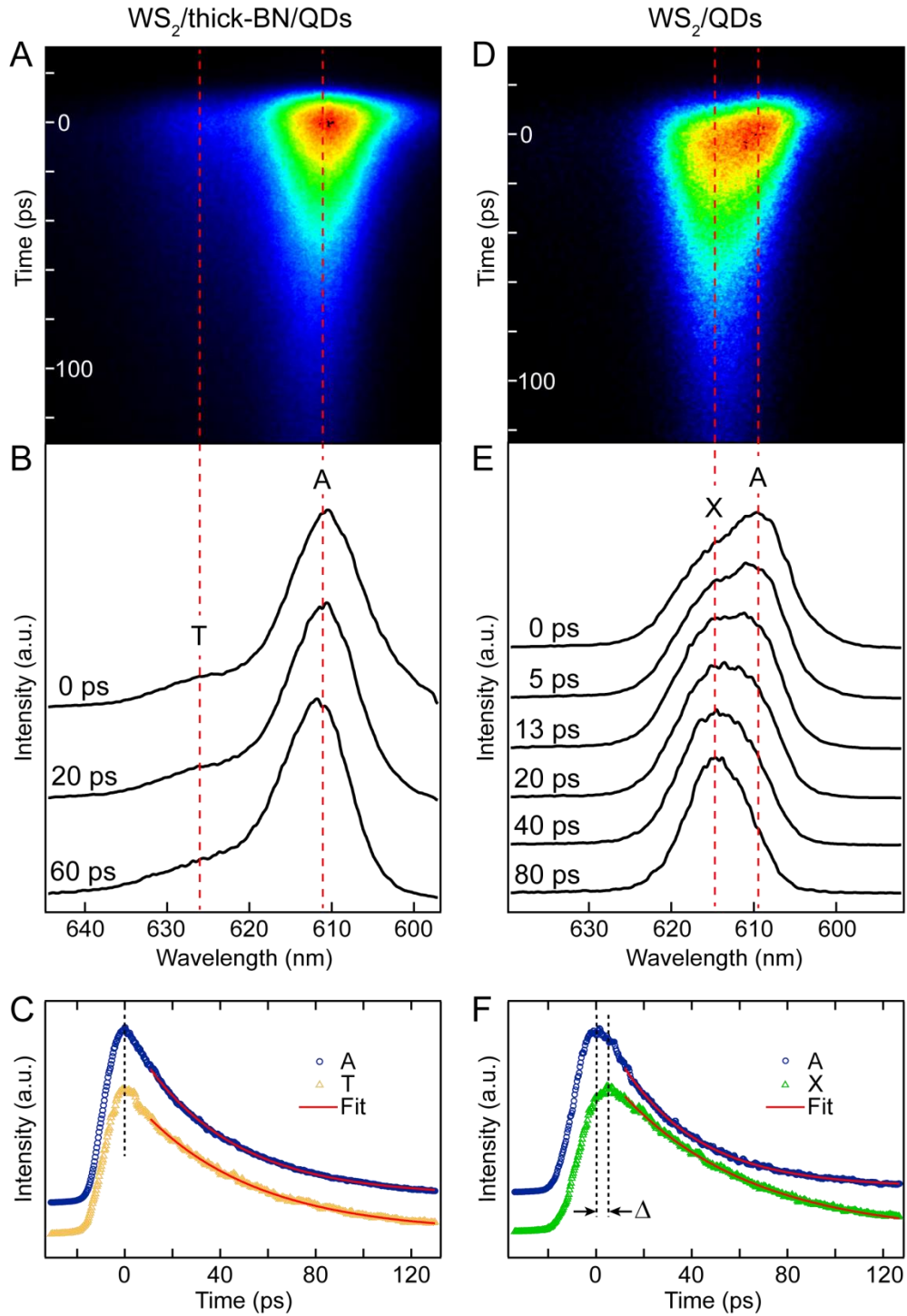

**Fig. 3.** Time-resolved PL spectra. The TRPL images, time evolutional PL spectra, and the decay curves at each peak wavelength, measured from (*A-C*) the uncoupled region-II and (*D-F*) the coupled region-III, respectively. The time evolutional PL spectra are shifted for clarity. All the decay curves are fitted by a single exponential equation (see solid lines). A decay time of Δ ~ 7.4 ps for X emission to reach its maximum compared with that of A emission can be observed.

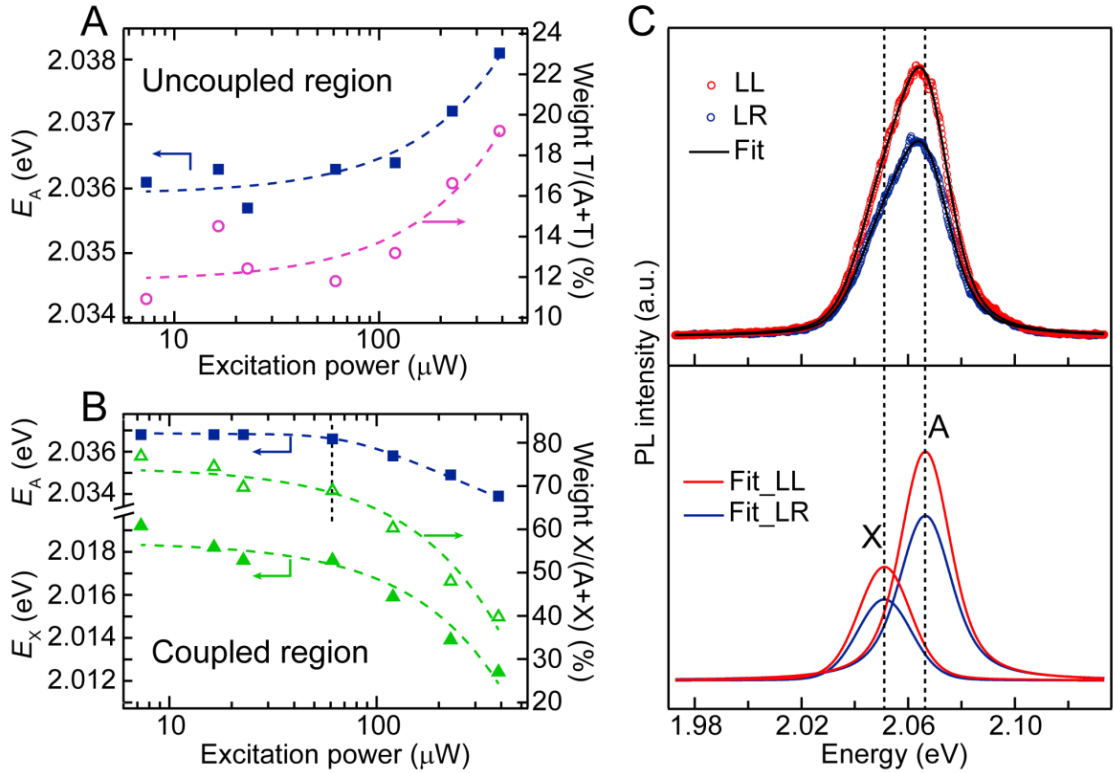

**Fig. 4.** Excitation power and circular polarization dependence of the PL. (*A*) Excitation power dependent peak energy of A emission and the spectral weight of T emission (defined by $I_T/(I_A+I_T)$) in the uncoupled region-II. The dashed lines are guides for eyes. (*B*) Excitation power dependent peak energies of A emission and X emission, and the spectral weight of X emission (defined by $I_X/(I_A+I_X)$) in the coupled region-III. The dashed lines are guides for eyes. (*C*) Circularly polarized PL spectra of the A emission and X emission, with excitation left circularly polarized and detection left circularly polarized (LL) or right circularly polarized (LR), solid lines are fitting curves. The circular polarization is defined as $(I_{LL} - I_{LR})/(I_{LL} + I_{LR})$, where $I_{LL(LR)}$ is the intensity of the left- (right-) circularly polarized PL spectra. The same circular polarization is obtained for A emission and X emission (16%).

# Supplementary Information for

Interfacial bound exciton state in a hybrid structure of monolayer WS$_2$ and InGaN quantum dots

Guanghui Cheng, Baikui Li, Chunyu Zhao, Xin Yan, Hong Wang, Kei May Lau, Jiannong Wang

Corresponding authors: Baikui Li, Jiannong Wang
Emails: libk@szu.edu.cn, phjwang@ust.hk

**This PDF file includes:**

    Supplementary text 1 to 3
    Figs. S1 to S2
    References for SI reference citations

**Supplementary Information Text**

**1. Band alignment analysis of the hybrid WS$_2$/QDs system.**

The peak energy of PL emission from InGaN QDs is 2.38 eV. Based on the quadratic dependence of emission energy on the indium composition (1), we could extract the stoichiometry of the QD to be In$_{0.28}$Ga$_{0.72}$N. The electron affinity of the QDs is estimated to be 4.82 eV based on the stoichiometry and the formula in ref. 2. The conduction band minimum (CBM) and valence band maximum (VBM) of QDs can be estimated around 4.82 eV and 7.25 eV, respectively. For WS$_2$ monolayer with the electron affinity 3.75 eV, a binding energy of 0.71 eV (3) and an optical bandgap of 2.04 eV in our case, the CBM and VBM are extracted to be about 3.75 eV and 6.5 eV, respectively. As a consequence, the close contacted WS$_2$ monolayer and InGaN QDs form a type-II band alignment. If an electron in the CBM of QDs binds with a hole in the VBM of WS$_2$, the radiative energy should be about 1.63 eV (smaller if considering the binding energy), which cannot be associated to the observed X emission with an energy close to intralayer exciton A. Note that although other works have reported slightly different energy values and binding energies, they do not qualitatively alter the above estimation.

**2. Excitation power dependences of emission energy and PL lifetime of InGaN QDs.**

When grown along the [0001] direction, InGaN quantum dots possess a built-in field due to the strain-induced piezoelectric polarization and spontaneous polarization. Such a built-in field tilts the band profile and results in a Stark shift for optical transition energy (4). PL spectra of InGaN QDs were measured (see Fig. S1*A*) with various excitation powers. As shown in Fig. S1*A*, the peak energy shows blue shift under higher excitation powers, indicating that the photoexcited carriers tend to compensate the built-in field and reduce the Stark shift.

We have also performed time-resolved PL measurements on InGaN QDs in the hybrid structures with various excitation powers. The lifetimes of QDs emission, as plotted in Fig. S1*B*, are in the order of nanoseconds which is much longer than that of the WS$_2$ and/or the interfacial bound exciton X.

**3. Structure characterizations of InGaN QDs.**

Transmission electron microscope (TEM) and atomic force microscope (AFM) images of the InGaN QDs are shown in Fig. S2. The typical structure of our InGaN quantum dots is a truncated pyramid. The QDs are estimated to hold an average height of approximately 2.5 nm, an average diameter of 66 nm and a density of approximately $1 \times 10^{10}$ cm$^{-2}$.

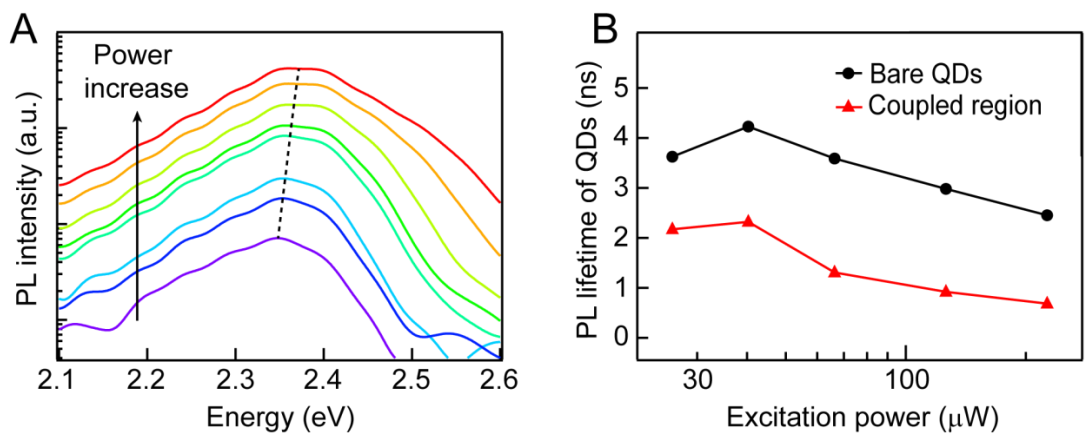

**Fig. S1.** (*A*) PL spectra of QDs at various excitation power (9 μW to 470 μW). (*B*) Lifetime of QDs as a function of the excitation power.

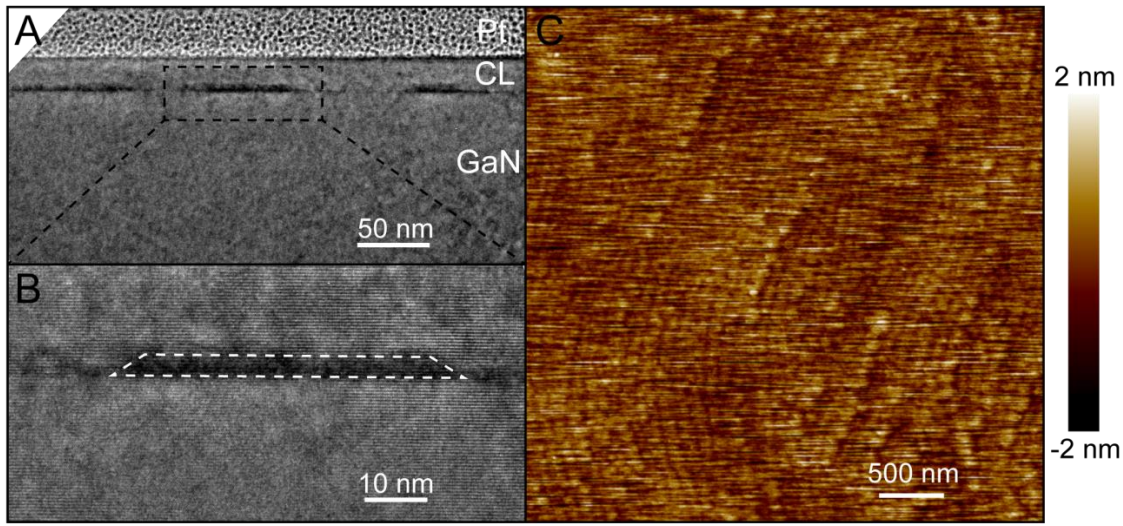

**Fig. S2.** Typical structure of InGaN self-assembled quantum dots. (*A* and *B*) Cross-sectional transmission electron microscope image of covered quantum dots. (*C*) Atomic force microscope image of uncovered quantum dots. A thin capping layer (CL) is intentionally deposited on the quantum dots to have a clear TEM visualization, while keeping its structure unchanged.